\documentclass[%
 reprint,
 amsmath,amssymb,
 aps,
pra,
floatfix,
]{revtex4-2}

\usepackage{graphicx}
\usepackage{dcolumn}
\usepackage{bm}
\usepackage{lipsum}
\usepackage{xcolor}

\begin{document}

\preprint{APS/123-QED}

\title[]{On the shape of ice stalagmites}

\author{Daniel Papa}
\author{Christophe Josserand}
\author{Caroline Cohen}
\affiliation{LadHyX, UMR7646 du CNRS \\ Ecole Polytechnique \\ 91128 Palaiseau, France
}

\date{\today}

\begin{abstract}

The growth of ice stalagmites obtained by the solidification of impacting droplets on a cooled substrate ($-50^{\circ}$C to $-140^{\circ}$C) is investigated experimentally. It is shown that for any combination of substrate temperature and drop discharge, there is a critical height above which unfrozen water accumulates at the stalagmite's tip, drips and develops into fingers that give a star-shape to the stalagmite. Both the vertical growth and the radial growth of the stalagmite are discussed through the Stefan problem and mass scaling arguments respectively. Finally, a phase diagram that presents the stalagmite aspect ratio in function of the main control parameters is proposed.

\end{abstract}

\maketitle

Speleothems are morphological patterns commonly observed in caves where mineral deposits accumulate over time to form a wide variety of shapes, among which draperies, flowstones, helictites, stalactites and stalagmites \cite{meakin10}. These geological patterns are the result of a complex interaction of hydrodynamics, chemical processes, erosional and depositional forces and interestingly some of them have an ice counterpart, such as ice pinnacles or scallops for instance, carved by water melting \cite{ristroph22}. Man-made artificial glaciers, like ice stupas used as a water storage device, are very similar in shape with these peleothems \cite{clouse17}. Stalactites are long and slender patterns that form from a fluid layer that controls precipitation of calcite while flowing down the surface \cite{short05a}. They have also their ice equivalent in icicles \cite{short06} where contrary to stalactites the growth is controlled by the solidification of the water layer through the removal of the latent heat heat by both convection and diffusion in air \cite{pegler21}. Similarly, stalagmites are long and slender vertical structures, but they grow from a calcite supersaturated solution that sits on its surface, either when a new drop of solution arrives or when a thin continuous water film spreads out radially down the stalagmite at sufficiently high discharge \cite{romanov08}. 
To the best of our knowledge, few studies have focused on their ice counterpart, ice stalagmites, in part because they seldom occur in nature and also because the condition for their appearance is quite rare: a continuous flow of droplets impacting on a sufficiently cold substrate (below freezing point). Another interesting features of stalagmites that has not been investigate is their radial growth. Here again, its seldom appearance in nature is due to the high feed rate needed to achieve both vertical and radial growth, even though this have been observed for granular towers \cite{chopin11}.

In this letter we study the growth of ice stalagmites by an experimental investigation. We delve into both the vertical and radial growths of the ice structure and propose a general explanation for their behaviour.

Experiments consists of releasing drops of water, of volume $V_{d} = 9.2\times10^{-3}$ ml corresponding to a radius $r_{0} = 1.3$ mm, with a period $t_d=V_{d}/Q$ for a given discharge $Q$, on a cold brass substrate. The discharge is imposed by a syringe pump and the substrate is cooled down by a bath of liquid nitrogen ($-196^{\circ}$C). By adjusting the level of liquid nitrogen, we are able to maintain a constant temperature of the substrate $T_{s}$ and impose a fixed temperature boundary condition between $-50^{\circ}$C and $-140^{\circ}$C. The system is located in a box where the air humidity is reduced to limit frost formation on both the stalagmite and the substrate. The falling height of the drops is kept at $H = 15$ cm for all experiments, leading to an impact velocity of $U_{0} \sim \sqrt{gH} = 1.4$ m/s. The evolution of the stalagmites is monitored with two Nikon D800 camera, one in front of the setup and one above, to visualize the vertical and radial growth of the stalagmites.

\begin{figure*}[t!]
  \includegraphics[width=\linewidth]{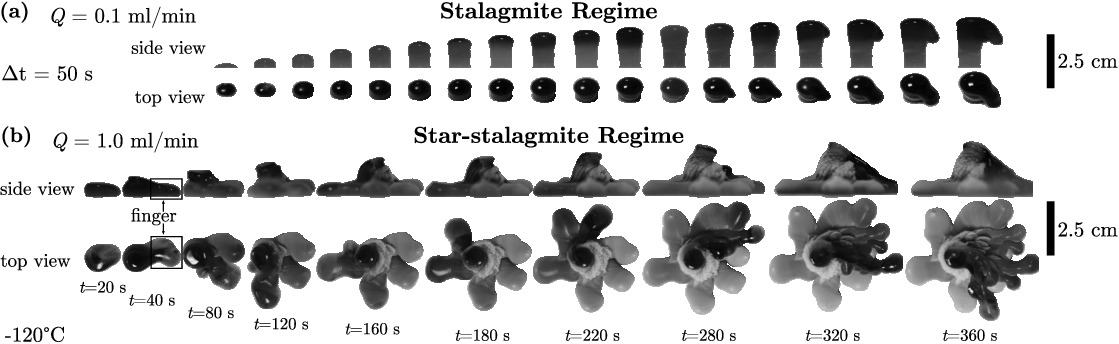}
  \caption{\label{fig:fig1} (a) Chronophotography of the stalagmite growth for a total time of 15 minutes and pictures taken at a regular time interval ($\Delta t = 20$ s), a flow discharge of $Q = 0.1$ ml/min at $-120^{\circ}$C. (b) Chronophotography of the star-shaped stalagmite growth for a total time of 6 minutes for a flow discharge of $Q = 1.0$ ml/min at $-120^{\circ}$C. Times are indicated below each picture.}
\end{figure*}

Figure \ref{fig:fig1} shows the chronophotographies of two experiments at the same substrate temperature $T_{s} = -120^{\circ}$C for two discharges, $Q = 0.1$ ml/min and $Q = 1.0$ ml/min corresponding to a period of 5.5 s and 0.55 s between two drop releases (figure \ref{fig:fig1} (a) and (b) respectively). At early times, each drop immediately solidifies in a pancake shape of radius $r_{max}$ close to the maximal spreading radius of the impacted drop on the cold substrate \cite{josserand16}. Experimentally we determined $r_{max} = 3.8$ mm. The stalagmite's shape can be thus firstly assimilated to a cylinder of radius $r_{max}$, of height $z(t)$, increasing linearly with time and with a rounded top, very similar to stalagmites commonly observed in caves formed by the precipitation kinetics of calcite \cite{romanov08}. This first regime of linear growth is expected to last until the tip of the stalagmite starts inflating when water accumulates on the tip (stalagmite 7, 8 on figure \ref{fig:fig1} (a)). The overflow of non-solidified liquid at the top of the stalagmite has also been observed for the growth of granular towers \cite{chopin11}. From this critical height $z_c$ upwards, water accumulates and spills over the tip and forms successive radial protrusions that we call fingers. The early stage of the first finger is illustrated on the last six stalagmites in figure \ref{fig:fig1} (a).

At higher discharges ($Q = 1.0$ ml/min), fingers develop quasi instantaneously (see figure \ref{fig:fig1} (b)). Each finger grows as a flowing rivulet on the cold substrate. After reaching a certain length ($\sim 3 \, $cm), the finger stops growing:  an another finger is initiated and tends to grow in its vicinity at an angle generally lower than 90$^{\circ}$ (see top view on figure \ref{fig:fig1} (b)). Some fingers were observed to grow at larger angles, typically around $180^{\circ}$ but these cases were seldomly observed experimentally. When the substrate has been covered with fingers ($t = 320$ s on figure \ref{fig:fig1} (b)), fingers flow and grow on top of the previous ones. The overall shape of the stalagmite is broadly conical ($t = 320$ s and $t = 360$ s on side view on figure \ref{fig:fig1} (b)) and reminiscent of volcanic cones which result from a roughly similar dynamic: a warm liquid flowing on a cold substrate that undergoes solidification while flowing downstream \cite{griffiths00}. To summarize this qualitative discussion, stalagmites growth can be separated into two dynamical regimes: a regime where the growth is purely vertical and a regime where both vertical and radial growth coexist and where the stalagmite develops its star-shape.

\begin{figure}[t!]
  \includegraphics[width=\linewidth]{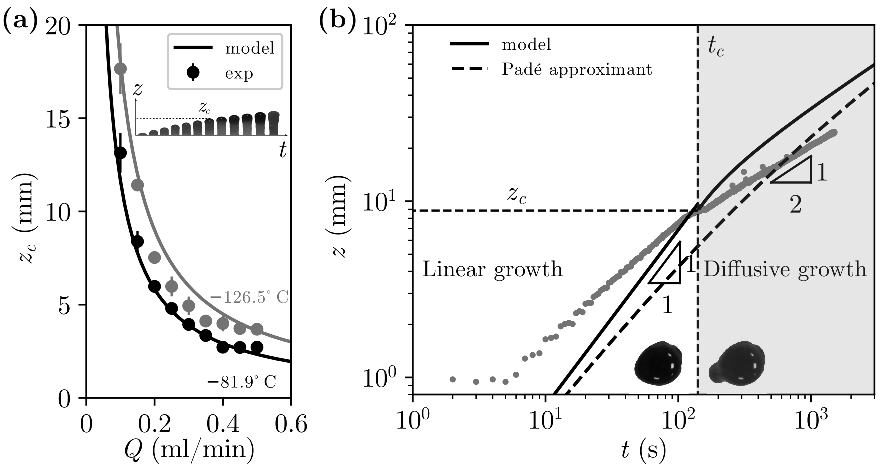}
  \caption{\label{fig:fig2} (a) Evolution of the critical height $z_{c}$ in function of the water discharge for two different substrate temperatures $T_{s} = -81.9^{\circ}$C in black lines and markers and $T_{s} = -126.5^{\circ}$C in gray lines and markers. The markers represent the experimental measurements and the solid lines represent equation \eqref{eq:zc}. (b) Temporal evolution of the stalagmite's tip for $Q = 0.2$ ml/min at $T_{s} = -77.7^{\circ}$C. The gray dots represent the experimental measurements, the black line represents equations \eqref{eq:z_linear} and \eqref{eq:ht} and the black dashed line represents equation \eqref{eq:Pade}. The two pictures underneath depict the stalagmite before and after the first finger has developed.}
\end{figure}

To understand the transition between these two regimes, we have to investigate how the liquid of an impacting drop solidifies. To freeze an incoming droplet, the stalagmite has to remove the latent heat released by its solidification $E_{s} = \rho_{i} L V_d$ via the diffusive heat flux through its top surface $\Phi_{\Delta T} = k_{i} \pi r_{max}^2 \Delta T/z$ where $\rho_{i}$, $L$, $k_{i}$ and $\Delta T = T_{m} - T_{s}$ are the ice density, water latent heat of solidification, ice thermal conductivity and temperature difference respectively ($T_{m} = 0^{\circ}$C, all thermal properties are found in table \ref{tab:tab1} in the appendix). The critical height $z_{c}$ corresponds to the height at which a drop solidifies exactly when the following arrives ($\Phi_{\Delta T} \times t_d = E_{s}$), in other words when the stalagmite accumulates water at each impact and start developing a finger. It reads

\begin{equation}
  \label{eq:zc} z_{c}= \frac{ \pi r_{max}^2 k_{i}\Delta T}{\rho_{i} Q L_f}.
\end{equation}
The corresponding time $t_c$ for this transition can be estimated by a simple mass balance $ \rho_i \pi r_{max}^2 z_c = \rho_l Q t_c$ and therefore reads $t_{c} = \pi^2 r_{max}^4 k_{i} \Delta T/(\rho_{l} L_{f} Q^2) $. Figure \ref{fig:fig2} (a) shows good agreement between the measurements of $z_{c}$ and equation \eqref{eq:zc} for two different temperatures and varying discharge. $z_{c}$ is proportional to $\Delta T$ and inversely proportional to $Q$. Physically it means that the stronger the heat flux, the higher the stalagmite can grow before it reaches $z_{c}$ and develops fingers (for a given $Q$). Conversely higher $Q$ means more heat to remove and therefore a lower capacity of the stalagmite to grow before reaching $z_{c}$. Therefore $z_{c}$ marks the height at which the purely vertical growth regime stops and where radial growth starts. We measure $z_{c}$ using fluorescein that emits green light under ultraviolet light, while it does not illuminate when frozen, allowing us to precisely detect the presence of liquid on the stalagmite's tip. 

Below $z_{c}$ the stalagmite grows by successive solidification of each impacting droplet and its height is therefore proportional to the time and the drop discharge and writes

\begin{equation}
    \label{eq:z_linear} z(t) = \frac{\rho_{l} Q }{\rho_{i}\pi r_{max}^2}t .
\end{equation}

We readily observe that $z(t_{c}) = z_{c}$. Above $z_{c}$ water is always present at the stalagmite's top meaning that it grows as a solidification front in water where the diffusive heat flux at the front is balanced by the solidification rate. Balancing the diffusive timescale $t_{D} = h^2/D_i$ (with $D_i = k_{i}/(\rho_{i} c_{p,i})$ the heat diffusion coefficient of ice and $h$ the water layer thickness) and the solidification timescale $t_{s} = \rho_{i} L h^2/(k_{i} \Delta T)$ we obtain the Stefan number $St = t_{D}/t_{s} = c_{p,i} \Delta T/L_f$. In our coldest experiments, $St = 0.88$ for $T_{s} = -140^{\circ}$C, indicating that diffusion is faster than solidification, so that the dynamics is governed by the solidification front and that the temperature field in the ice is stationary.

The temperature field in both ice and water can be deduced from the steady diffusion equations in each domain, time dependence coming simply from the solidification front $h(t)$. It has been shown that in this situation the heat diffusion in water can be neglected, leading to the simplified Stefan equation balancing the heat flux in the ice with the phase change~\cite{ghabache16,thievenaz19}. We solve the steady diffusion equation on the vertical axis to obtain the temperature field in the stalagmite $\partial^2 T/\partial z^2 = 0$ with the boundary conditions $T(0,t) = T_{s}$ and $T(z(t), t) = T_{m}$. The solidification front follows the Stefan boundary condition that states that the solidification rate is balanced by the heat flux at the boundary $\rho_{i} L \dot{h} = k_{i}\partial T/\partial z$ at $z = h(t)$. This yields a partial differential equation for $h(t)$ that results in the time evolution of the solidification front 

\begin{equation}
  \label{eq:Stefan} \rho_i L_f \frac{dh}{dt}=\frac{k_i \Delta T}{h},
\end{equation}

which can be solved in the second regime with the initial condition $h(t_c)=z_c$, yielding

\begin{equation}
  \label{eq:ht} h(t) = \sqrt{z_{c}^2 + 2 St \cdot D_i \cdot (t - t_c)}.
\end{equation}

Figure \ref{fig:fig2} (b) shows the temporal evolution of the stalagmite's tip for $Q = 0.2$ ml/min at $T_{s} = -77.7^{\circ}$C. The gray dots represent the experimental measurements, the solid black line represents the coupling of equation \eqref{eq:z_linear} (for $t < t_c$, the white background) and \eqref{eq:ht} (for $t > t_c$ the gray background). The transition from linear growth to diffusive growth is clearly observed by the change of slope of $z$ from 1 to 1/2 when $z$ reaches the critical height $z_{c}$ (horizontal dashed line), also the linear regime seems to be already affected by the diffusive one for $z<z_c$. Moreover, the model overestimates the height of the stalagmite by about 1 cm in the diffusive one although it captures correctly the dynamics (slope 1/2). The discrepancy between experimental results and the model may come from several factors, among which the fact that our model considers the water to be at the melting temperature, whereas the impacting drops are not (around 10 to 15$^{\circ}$C). This extra heat has to be removed by the stalagmite and may slow the solidification process. Furthermore, the heat diffusion in the solid substrate is also neglected, which would also reduce the ice growth.
Such general problem coupling heat diffusion in the substrate, ice layer and water (if needed) has been solved in one dimension leading to a self-similar solution when starting with the water in contact of the substrate following $h = \sqrt{D_{eff} t}$ where $D_{eff}$ is an effective diffusion coefficient that depends on the thermal properties of each material \cite{thievenaz19}. However, the difference between $D_{eff}$ and $2 St \cdot D_i$ is small in our case (few percent) and cannot explain the difference between the model and the measures. In fact, the observation that diffusion is already at play in the first regime indicates that the two behaviors (individual droplet freezing at short time and diffusion at larger one) are intricated and that a combination of both regime has to be taken into account in the model. It can be done simply by considering the Pad\'e approximant mixing the two formula \eqref{eq:z_linear} and \eqref{eq:ht}, yielding:

\begin{equation}
\label{eq:Pade} h(t) = \frac{t}{C_1+\sqrt{\frac{t}{C_2}}}.
\end{equation}
where the quantities $C_1=\frac{\rho_{i}\pi r_{max}^2}{\rho_{l} Q }$ and $C_2=2 St \cdot D_i$, can be deduced from the two regimes. Nevertheless, fitting these two parameters, close to these theoretical values (we find $C_1=3$ $\mathrm{s}\cdot \mathrm{mm}^{-1}$ and $C_2=1$ $\mathrm{mm}^{2}\cdot \mathrm{s}^{-1}$ provides a very good account of the short term dynamics (dashed line on figure \ref{fig:fig2} (b)), while it still overestimates it after $150$s, when the fingers start to form so that our 1D model is not anymore valid.

\begin{figure}[t!]
  \includegraphics[width=\linewidth]{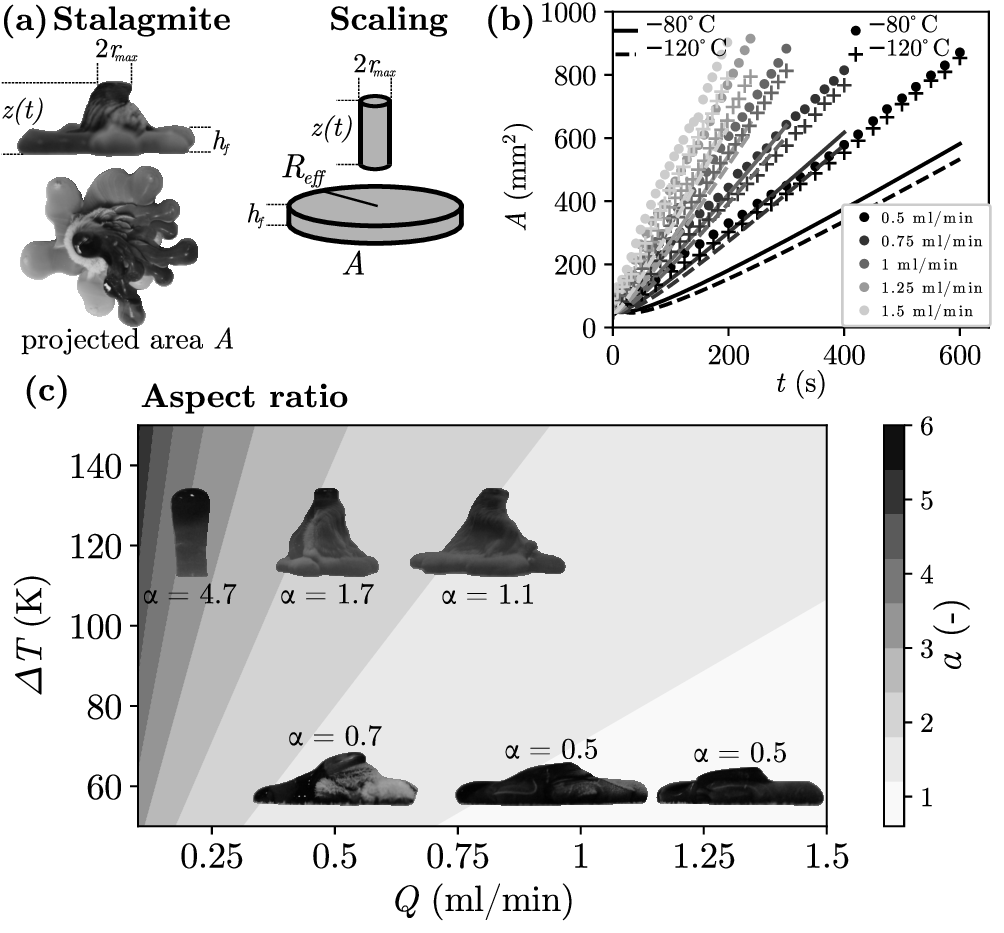}
  \caption{\label{fig:fig3} (a) Geometric characteristics of the stalagmite and the scaling model. (b) Experimental measurement of the projected area for 5 discharges $Q$ (markers in shades of gray) and two substrate temperatures (circles for $T_{s} = -80^{\circ}$C and pluses for $T_{s} = -120^{\circ}$C). (c) Phase diagram of the stalagmite aspect ratio.}
\end{figure}

A full understanding of the stalagmite radial growth would imply a thorough investigation of the finger development that can be seen as the solidification of a flowing rivulet \cite{huerre21}, which is out of the scope of the present Letter. Insights on the radial growth can be obtained by separating the growth in two parts: a vertical growth of a cylinder or radius $r_{max}$ and height proportional to $h(t)$ (equation \eqref{eq:ht}) and a radial growth of fingers of length  $l$ and thickness $h_{f}$ (see figure \ref{fig:fig3} (a)). Experimentally we measured $h_{f} = 6.9$ mm and observed that it remained constant for all substrate temperatures ($-50^{\circ}$C to $-140^{\circ}$C) and flow discharges ($0.1$ to $1.5$ ml/min). Seen from above, these fingers have a projected area $A$ on the substrate that, at first order, can be approximated as a disk of effective radius $R_{eff}$ and thickness $h_{f}$ (see scaling on figure \ref{fig:fig3} (a)). The total volume $V = Qt$ is distributed into the volume of the cylinder $V_{c} = A_{c}h(t)$ with $A_{c} = \pi r_{max}^2$ and the volume of the disk $V_{d} = Ah_{f} - A_{c} h_{f}$. Conservation of volume implies $V = V_{c} + V_{d}$ and manipulating the terms yields to the scaling relationship:

\begin{equation}
    \label{eq:A} A(t) = \frac{Q}{h_{f}}t - \frac{A_{c}}{h_{f}}\sqrt{z_{c}^2 + 2St \cdot D_i \cdot (t - t_c)} + A_{c}.
\end{equation}

The projected area was measured for different flow discharges at two substrate temperatures (figure \ref{fig:fig3} (b)). The model follows the experimental data relatively well for the higher discharges (1.5, 1.25 and 1 ml/min). For $-120^{\circ}$C) a slight decrease of $A$ is observed ($+$ markers in figure \ref{fig:fig3} (b)). A larger $\Delta T$ results in more mass solidified vertically hence a lower $A$. This behaviour is correctly captured by equation \eqref{eq:A} (dashed lines in figure \ref{fig:fig3} (b)). There are two asymptotic behaviours of equation \eqref{eq:A}. For smaller $Q$ the volume is mostly solidified as a purely vertical stalagmite, which explains why equation \eqref{eq:A} underestimates the projected area $A$ (the black solid and dashed lines on figure \ref{fig:fig3} (b)). For larger $Q$ all the volume is solidified as a disk of ever growing radius , therefore equation \eqref{eq:A} overestimates $A$. It is possible to define an effective radius $R_{eff} = \sqrt{A/\pi}$ that allows us to synthesize in one variable the effective spreading of the stalagmite on the cold substrate. However this scaling law is valid as long as the surface is not entirely covered by fingers. Once it happens (for instance $t=280$ s on figure \ref{fig:fig1} (b)) ice starts solidifying on top of the previous fingers and the shape of the stalagmite can no longer be assimilated to a cylinder on top of a disk but rather as something vaguely conical reminiscent of volcanoes (see $t=360$ s on figure \ref{fig:fig1} (b)).

Combining equations \eqref{eq:ht} and \eqref{eq:A} we can define the stalagmite's aspect ratio $h(t)/R_{eff}$. In the limit of long times (i.e. $t \rightarrow \infty$) we derive the stalagmite asymptotic aspect ratio

\begin{equation}
    \label{eq:aspect_ratio} \alpha = \sqrt{\frac{Q}{\pi h_f D_{eff}}}.
\end{equation}

Equation \eqref{eq:aspect_ratio} is shown on figure \ref{fig:fig3} (c) as a phase diagram and is compared with experimentally observed stalagmites. A relatively good agreement is observed especially for stalagmites at higher $\Delta T$.

$\alpha$ is the balance between mass flux $Q/(\pi h_{f})$ and the heat flux $D_{eff}$. In order to grow the stalagmite has to remove the heat generated during the phase change of the mass flux via diffusion over an area determined by the diffusion coefficient. Therefore $D_{eff}$ determines the effective radius of the stalagmite in contact with the substrate for a given mass flux and this equilibrium therefore determines the stalagmite's aspect ratio.

To conclude, in this Letter we investigated the growth of ice stalagmites. We showed that all stalagmites grow linearly until they reach a critical height where they start accumulating water therefore entering in a Stefan configuration where the growth is govern by heat diffusion. This is also the moment when they grow radially and develop a characteristic star-shape. The final asymptotic aspect ratio of the stalagmite is the result of a balance between the two main fluxes that govern its growth, the mass flux and the thermal flux. This original experiment can be seen as a generic model for other growth system where a flux of a fluid undergoes solidification by flowing on a cold substrate, colder than the fluid fusion temperature.

\begin{acknowledgments}

This work was supported by Agence de l'Innovation de D\'efense (AID) - via Centre Interdisciplinaire d'Etudes pour la D\'efense et la S\'ecurit\'e (CIEDS) - (project 2021 - ICING).

\end{acknowledgments}

\appendix
\section{Physical constants}

\begin{table}[b]
    \centering
    \caption{\label{tab:tab1} Thermal and physical properties \cite{petrenko99}.}
    \begin{ruledtabular}
    \begin{tabular}{lllll}      
         & \textbf{Variables} & \multicolumn{3}{c}{\textbf{Numerical values}} \\
         &  &  ice & brass & water \\
        \hline
        $\rho$ & Density [kg/m$^3$] & 917 & 8600 & 1000 \\
        $k$ & Conductivity [W/(m K)] & 2.4 & 2.2 &  \\
        $c_{p}$ & Capacity [J/(kg K)] & 2090.0 & 376.8 &  \\
        $e$ & Effusivity [W~s$^{1/2}$~m$^{2}$~K$^{-1}$] & 2144.7 & 21818.0 &  \\
        $D$ & Diffusivity [m$^2$/s] & $1.3\times 10^{-6}$ & $4.5\times 10^{-5}$ & \\
        $L_f$ & Latent heat [kJ/kg] & & & 333.5 
    \end{tabular}
    \end{ruledtabular}
\end{table}

\nocite{*}

\bibliography{draft}

\end{document}